\documentclass[10pt,conference,a4paper]{IEEEtran}

\usepackage{paralist}
\usepackage{booktabs,latexsym,moreverb}
\usepackage{multirow,alltt,fancyvrb,amsmath, amssymb}
\usepackage{zed-csp, graphicx}
\usepackage{subfigure, amsthm}

\newcommand{\go}[1]{\stackrel{#1}{\longrightarrow}}
\renewcommand{\seq}{\mbox{~\large{;}}~}
\newcommand{\sq}{\mbox{~;~}}

\newcommand{\synch}[2]{#1\&#2}
\newcommand{\cpair}{\div}
\newcommand{\close}[1]{[~#1~]}

\theoremstyle{plain} \newtheorem{thm}{Theorem}

\theoremstyle{plain} \newtheorem{lema}{Lemma}

\theoremstyle{definition} \newtheorem{defn}{Definition}

\newenvironment{script}{\scriptsize\verbatim}{\endverbatim}

\begin{document}


\title{Formalizing cCSP Synchronous Semantics in PVS}



\author{\IEEEauthorblockN{Shamim H. Ripon}
 \IEEEauthorblockA{Department of Computing Science\\
 University of Glasgow, UK\\
 Email: shamim@dcs.gla.ac.uk}\\
 \and
 \IEEEauthorblockN{Michael Butler}
 \IEEEauthorblockA{School of Electronics and Computer Science\\
 University of Southampton, UK}}

\maketitle

\begin{abstract}
\emph{Compensating CSP} (cCSP) is a language defined to model long running business transactions within the framework of standard CSP process algebra. In earlier work, we have defined both traces and operational semantics of the language. We have shown the consistency between the two semantic models by defining a relationship between them. Synchronization was missing from the earlier semantic definitions which is an important feature for any process algebra. In this paper, we address this issue  by extending the syntax and semantics to support synchronization and define a relationship between the semantic models. Moreover, we improve the scalability of our proof technique by mechanically verifying the semantic relationship using theorem prover PVS. We show how to embed  process algebra terms and semantics into PVS and to use these embeddings to prove the semantic relationship.\\

\end{abstract}

\section{Introduction}
\label{Introduction}

Business transactions involve multiple partners coordinating and interacting with each other. These transactions have hierarchies of activities that need to be orchestrated. Business transactions also need to deal with faults that can arise at any stage of the transactions. Compensation mechanisms~\cite{gray:tranproc} are very important for handling faults for transactions that require a long period of time (also called \emph{Long Running Transaction}, LRT). Process calculi are models or languages for concurrent and distributed interactive systems. Based on the framework of Hoare's CSP process algebra~\cite{Hoare:CSP}, Butler~\emph{et al}~\cite{csp25} introduced compensating CSP, a language to model long running transactions. The language introduces a method to declare a transaction as a process and it has constructs for orchestration of compensations.

A formal semantics offers a complete, rigorous definition of a language and provides a foundation for mathematical proofs about programs. We have defined both traces~\cite{csp25} and operational semantics~\cite{cCSP05} of the language. Having two semantic models of a language, it is natural to verify the consistency between them and check how they are related. We have defined a relationship between the semantic models in~\cite{techreport-ecs} by following a systematic approach.

Synchronization is an important and well understood feature for concurrent and distributed processes. However, synchronization was not included in our work. In this paper, we extend the cCSP semantic models to define the semantics for synchronous processes, where processes synchronize over a set of synchronizing events, and non-synchronizing processes interleave with each other. We also show that the same relationship that was defined for asynchronous processes also hold for synchronous processes. We take our work one step further by mechanical verifying the relationship by using the theorem prover PVS~\cite{pvs92}. Mechanical verification overcomes the problem in hand proofs, also identifies potential flaws in the semantic definitions.

The rest of the paper is organized as follows. A brief overview of cCSP language is given in \S~\ref{sec:ccsp}. We then describe how the language terms are extended to define synchronization of processes in \S~\ref{sec:synch}. We also give an example of a web service specified by using cCSP and using the extended feature of synchronization. In the following two sections, we define how the trace and the operational semantics are extended to synchronization. \S~\ref{sec:relation} defines a relationship between the semantic models and sketches the proof steps. We describe the PVS embedding of cCSP syntax and semantics in \S~\ref{sec:mechanize}. These embeddings are then used to establish the relationship between the synchronous semantic models. We outline some complimentary work in the following section. Finally, we draw our conclusions in \S~\ref{sec:conclusion}.

\section{Compensating CSP}\label{sec:ccsp}

\begin{figure}[!htb]
\centering
\includegraphics[width=86 mm]{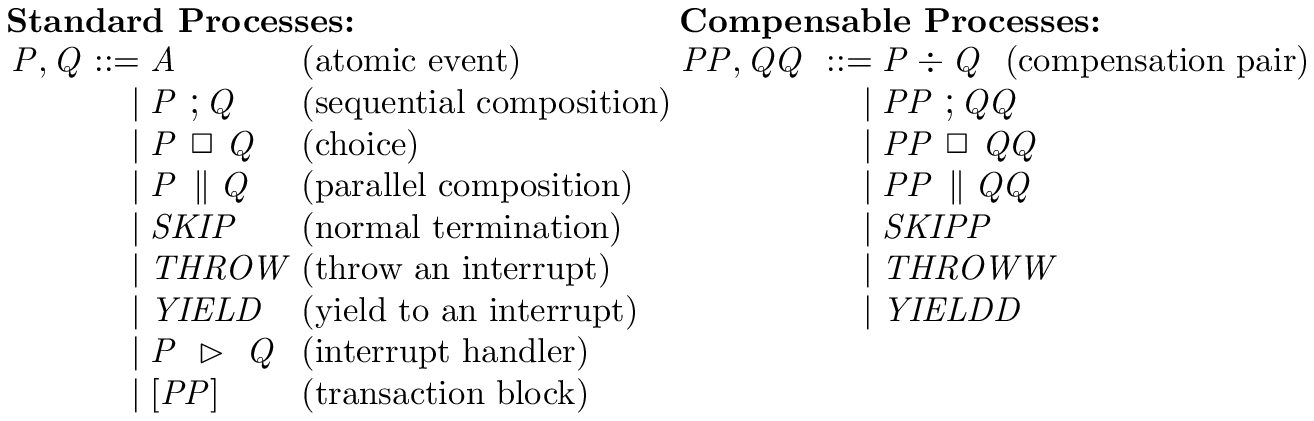}
\caption{cCSP syntax}
\label{fig:syntax}
\end{figure}

Processes in cCSP are modelled in terms of the atomic events they can engage in. The language provides operators that support sequencing, choice, parallel composition of processes. In order to support failed transaction, compensation operators are introduced. The processes are categorized into \emph{standard}, and \emph{compensable} processes. Compensation is part of a compensable process that is used to compensate a failed transaction. We use notations, such as, $P,Q,..$ to identify standard processes, and $PP,QQ,..$ to identify compensable processes. The asynchronous subset of cCSP syntax is summarized in Fig.~\ref{fig:syntax}.

The basic unit of the standard processes is an atomic event ($A$). The other operators are the sequential~($P\sq Q$), and the parallel composition ($P\parallel Q$), the choice operator ($P\extchoice Q$), the interrupt handler ($P\rhd Q$), the empty process $SKIP$, raising an interrupt $THROW$, and yielding to an interrupt $YIELD$. A process that is ready to terminate is also willing to yield to an interrupt. In a parallel composition, throwing an interrupt by one process synchronizes with yielding in another process. The basic way of constructing a compensable process is through a compensation pair ($P\cpair Q$), which is constructed from two standard processes, where $P$ is called the \emph{forward} behaviour that executes during normal execution, and $Q$ is the associated compensation that is designed to compensate the effect of $P$ when needed. The sequential composition of compensable processes is defined in such a way that the compensations of the completed tasks will be accumulated in reverse to the order of their original composition, whereas compensations from the compensable parallel processes will be placed in parallel.
By enclosing a compensable process $PP$ inside a transaction block $\close{PP}$, we get a complete transaction and the transaction block itself is a standard process. Successful completion of $PP$ represents successful completion of the block. But, when the forward behaviour of $PP$ throws an interrupt, the compensations are executed inside the block, and the interrupt is not observable from outside of the block. $SKIPP, THROWW$, and $YIELDD$ are the compensable counterpart of the corresponding standard processes and they are defined by pairing an empty compensation with them, e.g., $SKIPP~=~SKIP\div SKIP$.

\section{Extending cCSP with Synchronization}\label{sec:synch}

\begin{figure}[!htb]
\centering
\includegraphics[scale=.38]{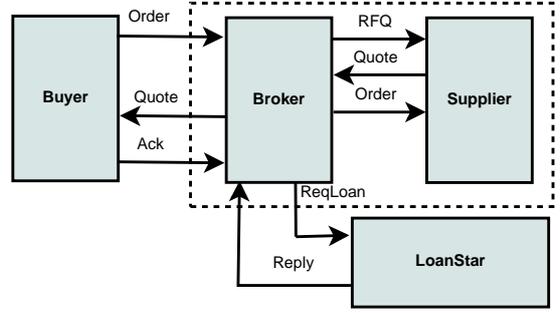}
\caption{Architectural view of Car Broker web Services }
 \label{fig:carbroker}
\end{figure}

We define a parallel operator synchronizing over observable events\footnote{We use normal and observable interchangeably;  normal event:  $a\in\Sigma$} extending our earlier definition, where processes interleave over observable events and synchronize only over terminal events\footnote{Cause  termination of a process term, a terminal event $\omega\in\Omega=\{\tick,~!,~?~\}$}.
We assume a set of events $X$ over which processes will synchronize. The process $(P\parallel_{X}Q)$ represents the parallel composition of processes $P$ and $Q$, synchronizing over the set of events $X$. Operationally, $P$ and $Q$ interact by synchronizing over the events from $X$, while events not in $X$ can occur independently. An event where both processes synchronize becomes a single event in ($P\parallel_{X}Q$), by a synchronizing operator which will be defined later. In the following example a business transaction is modelled by cCSP constructs added with synchronization:

\noindent{\bf Example: \emph{(Car Broker Web Services)}}~ We model a car broker web service {\bf Broker} which provides online support to customers to negotiate car purchases and arranges loans for these. The architectural view of the web service is given in Fig.~\ref{fig:carbroker}.

In cCSP, a process is described in terms of its interactions with its environment or with other processes by using atomic actions. The communications are defined via channels as in standard CSP. A communication is an event described by the pair $c.v$, where $c$ is the channel name and $v$ is the value of the message. Input/output are defined using same construct as in CSP. Concurrent processes communicate via channels. We also use I/O parameters for compensation pair: $$A?x\cpair B.x~\sq P(x)~~=~~ \square_{x\in S}~A.x\cpair B.x~\sq P(x)$$

The first step of the transaction is a compensation pair, where the primary action is to receive an order from the buyer and the compensation is to cancel the order. $M$ is used to represent the finite set of car models ranged over by $m$.

\begin{center}
\includegraphics[scale=.90]{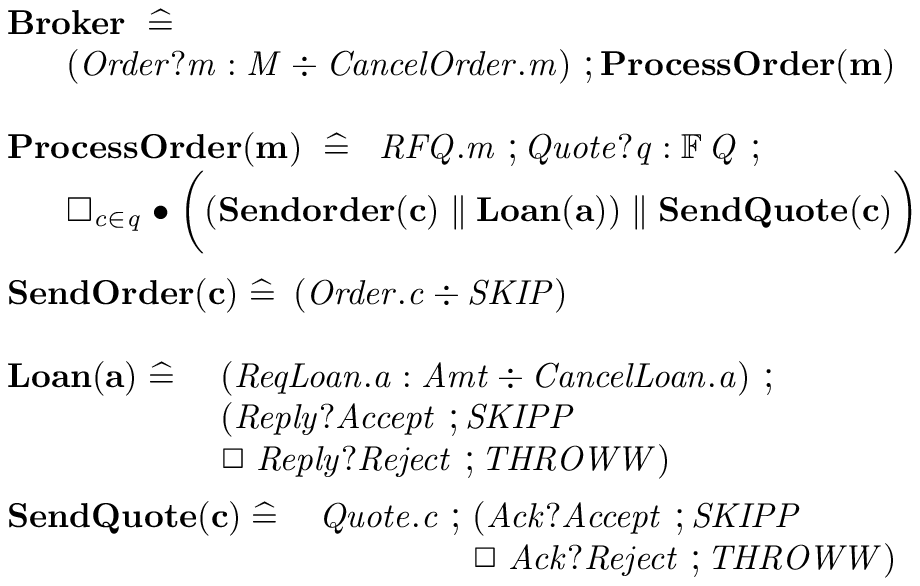}
\end{center}

The {\bf Broker} requests the {\bf Supplier} for available quotes (\emph{RFQ}) and then selects a quote from the received quotes (\emph{Quote}). The {\bf Broker} arranges a loan for the quoted car by requesting a loan from {\bf LoanStar}. The loan amount ($Amt$) of loan to be requested is decided from the selected quote and passed to the process {\bf Loan}. It requests loan from {\bf LoanStar} which is either accepted or rejected. If the loan cannot be provided then an interrupt is thrown to cancel the actions that have already taken place. A compensation is added to \emph{ReqLoan} (\emph{CancelLoan}) so that in the case of failure in a later stage the compensation can be invoked to cancel the event. the quote is also sent to the buyer ({\bf SendQuote}).
An interrupt can be raised either by the {\bf Buyer} by
rejecting the quote or by the {\bf LoanStar} by rejecting
the requested loan. In either case, the {\bf Supplier} will
terminate yielding an interrupt thrown by the {\bf Broker} and
compensations from both {\bf Broker} and {\bf Supplier} will run in parallel.

The behaviour of the car broker web service is defined by
combining the behaviour of {\bf Broker, Buyer, Supplier}, and {\bf
LoanStar}, where the processes synchronize over the sets $A,B$ and $C$.
\begin{eqnarray*}
{\bf System} &\defs& {\bf Buyer}\parallel_{A}
    \bigl[~{\bf Broker} \parallel_{B}
    {\bf  Supplier}~\bigr]\\
 &&   \parallel_{C} {\bf LoanStar}
\end{eqnarray*}
$\begin{array}{l}
  A=\set{Order,Quote,Ack},~~ B=\set{RFQ,Quote,Order}\\
  C=\set{ReqLoan, Reply}
\end{array}$

The example illustrates the synchronization of processes within a transaction block, $\close{{\bf Broker}\parallel_{B} {\bf  Supplier}}$  and between transaction blocks ({\bf Buyer} and {\bf LoanStar} are transaction blocks). It also outlines how compensations are handled in each case.

\section{Extended Trace Semantics}\label{sec:trace}

A trace records the behaviour of a process up to some moment in time. The traces of composite processes are defined in terms of their constituent processes. Processes are assumed to have an alphabet of actions $\Sigma$ which does not include the terminal events $\Omega = \set{\tick,!,?~}$. Terminal symbols indicate the way how a process terminates. Standard processes are defined as non-empty set of traces of the form $s\trace{\omega}$ where $s\in\Sigma^*$ and $\omega\in\Omega$. For traces $s$ and $t$, we write $s.t$ as their concatenation. Operators are first defined on traces and then lifted to set of traces to define processes. The traces of a standard process $P$ is denoted as $T(P)$. Compensable processes consist of a set of pair of traces of the form $(p\trace{\omega},p'\trace{\omega'})$, where $p\trace{\omega}$ represents the forward behaviour and $p'\trace{\omega'}$ represents the compensation. $T(PP)$ denotes the trace of a compensable process $PP$.

\begin{figure*}[!htb]
\centering
\subfigure[Standard]{
\includegraphics[width=68 mm]{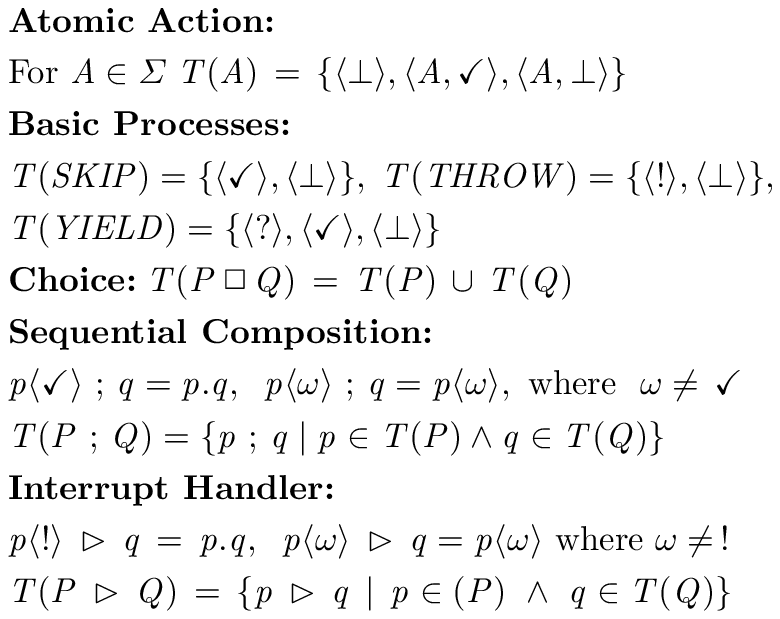}}
\subfigure[Compensable]{
\includegraphics[width=76 mm]{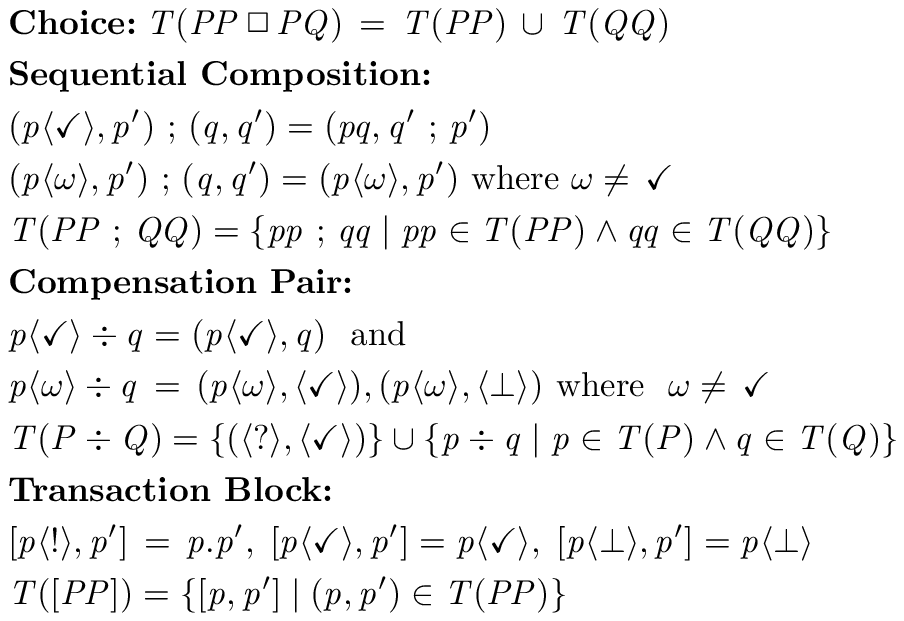}}
 \caption{Trace semantics of sequential processes}
 \label{fig:trace}
\end{figure*}

Parallel processes synchronize over synchronizing events and interleave over other events. When processes fail to synchronize, the execution blocks and we get a partial behaviour from the composition. To denote partial behaviour, we assume a special terminal symbol $\bot\in\Omega$ which indicates partial trace. Partial traces are analogous to trace prefixes in standard CSP. With the definition of partial behaviour, traces from standard processes satisfy the following properties:
\begin{itemize}[--]
    \item $\trace{\bot}\in T(P)$
    \item $p\trace{x}q\in T(P) ~~~\implies~~~p\trace{\bot}\in
    T(P)$ ~~($x\in\Sigma$)
\end{itemize}
We assume $\bot$ acts as a cut for trace concatenation: $
p\trace{\bot}q ~~=~~p\trace{\bot}$. With the introduction of the
new terminal event ($\bot$), we extend the original trace
definitions. The extended trace definitions for sequential
operators are defined in Fig.~\ref{fig:trace}.

We define a synchronization operator on events writing $\synch{A}{A'}$ for the synchronization of events $A$ and $A'$. Consider two processes synchronizing over events $a$ and $a'$, the synchronization is defined as: $\synch{a}{a}~=~a$, and $\synch{a}{a'}~=~\bot$ when $a\neq a'$ and do not synchronize with each other.

We define a synchronization operator over terminal events from the set $\Omega$. Table~\ref{tab:synch} enumerates the evaluation of this operator. We also define the synchronization operator to be commutative. From Table~\ref{tab:synch} it can be seen that the operator is well-defined for all the operands in the set $\Omega$. Case analysis shows that the synchronization operator is associative.
\begin{table}[!htb]
\centering \caption{Synchronization of terminal events}
\label{tab:synch} $\begin{array}{c|ccccccc}
\omega  ~& ~~!~&~!~&~  !~  &~?~&~  ?~  &\tick~&\bot\\
\omega' ~& ~~!~&~?~&~\tick~&~?~&~\tick~&\tick~&\omega
 \\\hline 
\synch{\omega}{\omega'} ~& ~~!~&~!~&~  !~  &~?~&~  ?~
&\tick~&\bot
\end{array}$
\end{table}

Assuming $a,a'\in X$ and $b,b'\nin X$, the parallel composition of traces from standard processes are defined as follows: \begin{eqnarray*}
\begin{array}{lll}
    \trace{\omega}~~\parallel_{X}  \trace{\omega'} &~~=~~&
        \set{~\trace{\synch{\omega}{\omega'}}~}
         \\
    \trace{a}p   \parallel_{X} \trace{\omega} &~~=~~&
        \set{~\trace{\perp}~}
        \\
    \trace{a}p   \parallel_{X} \trace{a'}q  &~~=~~&
         \set{~(\synch{a}{a'}) r ~|~ r \in (p \parallel_{X} q)~}
         \\[.5ex]
    \trace{b}p  \parallel_{X} \trace{\omega} &~~=~~&
        \set{~\trace{b}r ~|~ r\in (p\parallel_{X} \trace{\omega})~}
        \\
    \trace{b}p   \parallel_{X} \trace{a}q  &~~=~~&
         \set{~\trace{b}r ~|~ r \in (p \parallel_{X} \trace{a}q)~}
         \\
    \trace{b}p  \parallel_{X} \trace{b'}q  &~~=~~&
         \set{~\trace{b}r ~|~ r \in (p \parallel_{X} \trace{b'}q)~}
         \\
   &&  \cup~ \set{~\trace{b'}r ~|~ r \in (\trace{b}p \parallel_{X} q)~}
\end{array}
\end{eqnarray*}
The parallel and synchronization operators are symmetric. For
brevity we omit the symmetric cases. The parallel composition of
standard processes is defined as follows:
\begin{eqnarray*}
T(P\parallel_X Q) &=& \set{~r~~|~~r\in(p\parallel_X q)\\
        &&\land~~p\in T(P)~~\land~~q\in T(Q)~}
\end{eqnarray*}

With the definition of partial behaviour ($\bot$), a pair of
traces ($p\trace{\omega},p'\trace{\omega'}$) of a compensable
process satisfies the following properties: For $x\in\Sigma$,
\begin{itemize}[--]
 \item $(\trace{\bot},p')\in T(PP)$

\item $(p\trace{x}q,p')\in T(PP)\implies
 (p\trace{\bot},\_)\in T(PP)$

\item $(p,p'\trace{x}q')\in T(PP)\implies
 (p,p'\trace{\bot})\in T(PP)$
\end{itemize}
The trace semantics for compensable parallel processes is defined
as follows:\\
$(p,p') \parallel_{X} (q,q') =$ \\
$\begin{array}[t]{r}
\set{(r,r')|r\in (p\parallel_X q)
   \land r'\in(p'\parallel_X q')\land last(r)\neq\bot}\\[.5ex]
   \cup\; \{(r,\trace{\bot})| r\in (p\parallel_x q)\land
   last(r)= \bot\}
   \end{array}$
\begin{eqnarray*}
 T(PP \parallel_{X} QQ) &=&\set{~ rr ~|~ rr\in (pp\parallel_X qq)\\
&& \land pp\in T(PP) ~\land~ qq\in T(QQ) ~}
\end{eqnarray*}
$last(t)$ returns the terminal symbol from a
trace $t$.


\section{Extended Operational Semantics}\label{sec:op}

The operational semantics are defined by using labelled transition systems~\cite{Plotkin:OS}. Inference rules are used to define the transitions that a process may perform, which for composite processes are given in terms of the possible transition of the constituents (See~\cite{cCSP05} for detail). Two types of transition rules are defined: normal and terminal. Normal transition is caused by a normal event resulting in a transition of a process term from one state to another. Terminal transition is caused by a terminal event where standard process terms terminate to a null process and the forward behaviour of compensable process terms terminate leaving the attached compensation for future reference. Note that the language terms are extended to define the null (0) process that cannot perform any action. For standard and compensable process terms $P$ and $PP$ (where $P,PP\neq 0$), the normal and terminal transitions are defined as followed:

\begin{eqnarray*}
 P\go{a}P',&&  PP\go{a}PP'\quad(a\in\Sigma)\\
 P\go{\omega}0,&&
 PP\go{\omega}P~~\;(\omega\in\set{\tick,!,?})\\
 &&~~\mbox{($P$ is the compensation of $PP$)}
\end{eqnarray*}

We extend the transition rules by defining the transitions by a $\bot$ where both standard and compensable processes terminate to a null process. For any process terms $P$ and $PP$ (where $P,PP\neq 0$), the transitions by a $\bot$ are defined as follows:
\begin{eqnarray}
P\go{\bot}0, \qquad PP\go{\bot}0\label{eqn:bot}
\end{eqnarray}

\begin{figure*}[!tbh]
\centering
\subfigure[Standard]{
\includegraphics[width=82 mm]{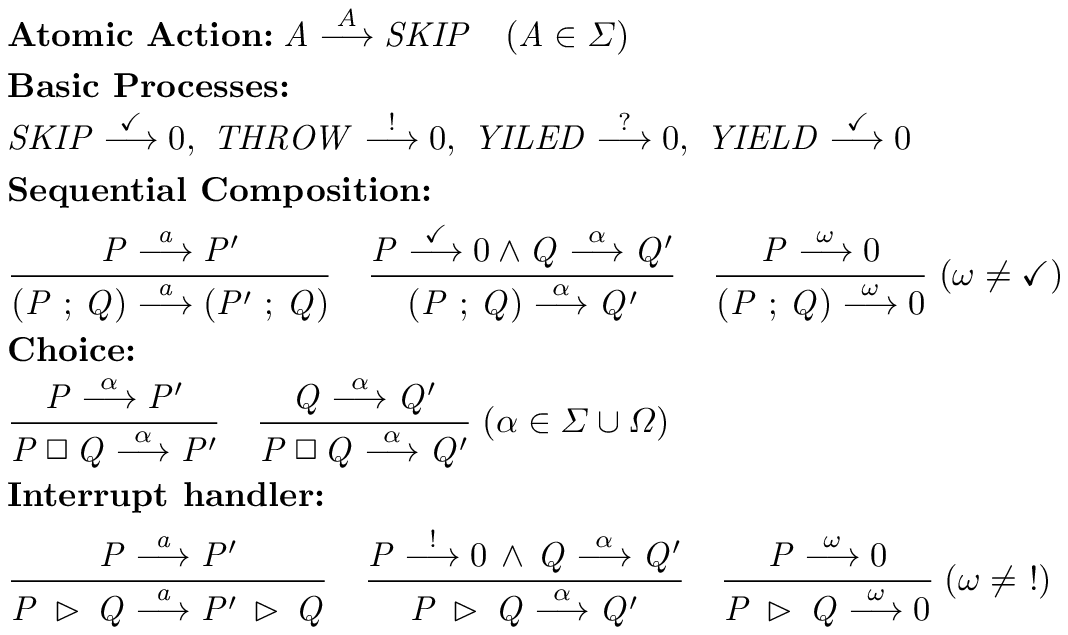}
\label{fig:os-std}}
\subfigure[Compensable]{
\includegraphics[width=82 mm]{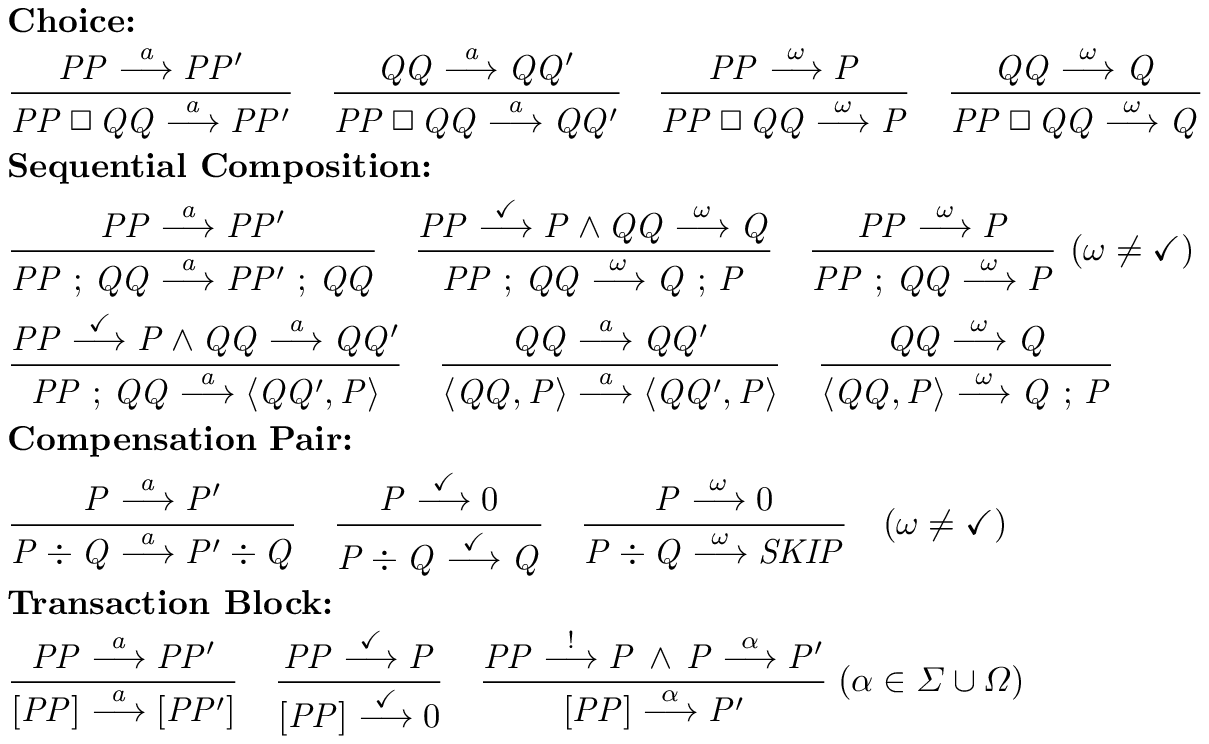}
\label{fig:os-comp}}
\caption{Operational Semantics for sequential processes}
\label{fig:op}
\end{figure*}

The transition rules defined in equation~(\ref{eqn:bot}) cover the transitions for both standard and compensable process terms by the $\bot$. Hence we do not need to define additional transition rules by a $\bot$. The transition rules for sequential standard and compensable processes are defined in Fig.~\ref{fig:os-std} and Fig.~\ref{fig:os-comp} respectively.

\begin{figure*}[!tbh]
\centering
\subfigure[Standard]{
\includegraphics[width=70 mm]{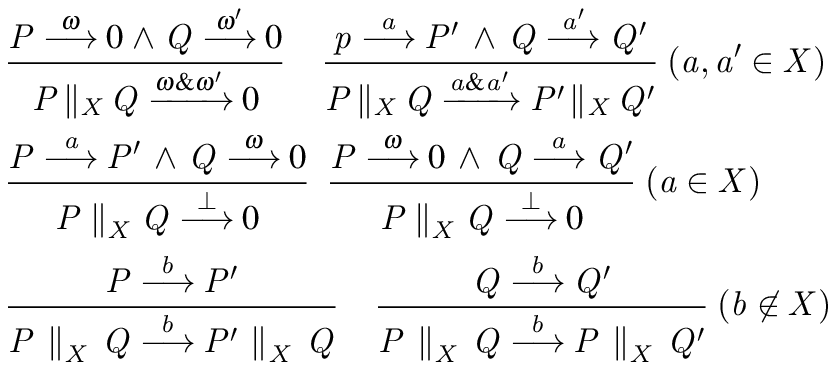}
\label{fig:synch-op}}
\subfigure[Compensable]{
\includegraphics[width=95 mm]{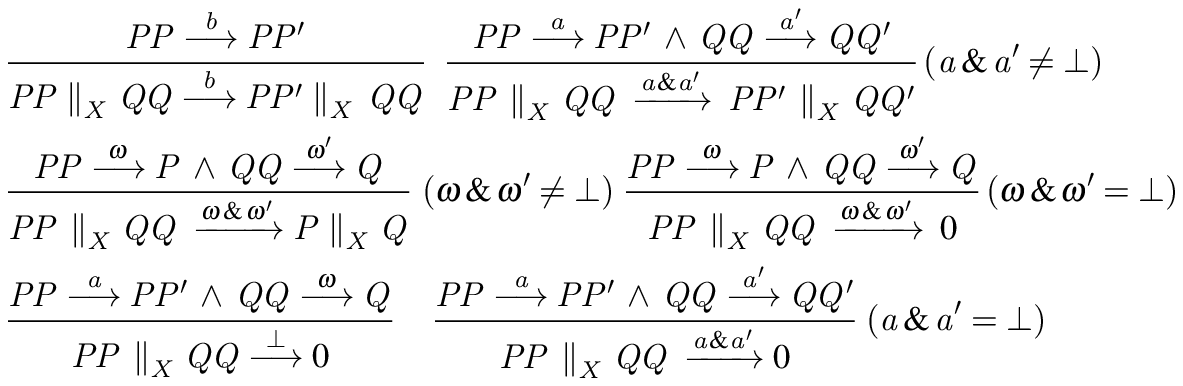}
\label{fig:csyn-op}}
\caption{Operational Semantics for synchronous processes}
\label{fig:op}
\end{figure*}

As $\bot$ is introduced during process synchronization and $\bot$
is a useful semantic device that helps us deriving semantic
correspondence, we define the extended transition rules for
parallel processes and define those transitions that introduce a $\bot$.
For a compensable process the transition by a $\bot$ lead to a null process and according to our definition no compensations are stored (being partial behaviour).
The transition rules for standard and compensable parallel processes are shown in Fig.~\ref{fig:synch-op} and Fig.~\ref{fig:csyn-op} respectively.

\section{Semantic Relationship}\label{sec:relation}

\begin{figure}[!htb]
\centering
\includegraphics[width= 68 mm]{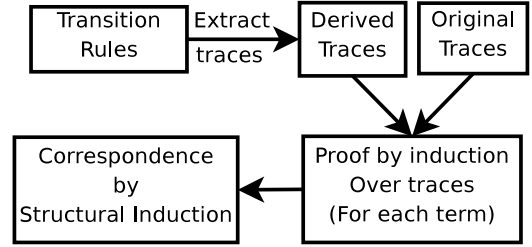}
\caption{Steps for semantic correspondence}
\label{fig:proof-steps}
\end{figure}

Over the years, several techniques have been used to establish  relationship between different semantic models. Widely used techniques are deriving one semantics from another (e.g.\cite{utp,os2ds}), extracting the behaviour from one semantic model and showing its relation with another (e.g.\cite{schneider:tcsp}) etc. Roscoe~\cite{Roscoe:CSP} outlines how to define the semantic relationship for CSP. In our earlier work~\cite{techreport-ecs,Ripon2008}, we have adopted a systematic approach showing a relationship between the semantic models. Traces are extracted from the transition rules of the operational semantics and show that the extracted traces correspond to the original traces for each term of the language and finally, prove the correspondence by structural induction over the process terms. The steps are depicted in Fig.~\ref{fig:proof-steps}.

In this paper, we extend our earlier approach to define and prove the relationship between the synchronous semantic models. Due to the introduction of partial behaviour, proving the correspondence for synchronous semantic modes becomes critical.  We briefly describe the steps shown in Fig.~\ref{fig:proof-steps} for asynchronous processes and extend those steps for synchronous processes.

The operational semantics leads to lifted transition relations
labelled by sequences of events. This is defined recursively. For
a standard  process $P$:
\begin{eqnarray*}
    P\go{\trace{\omega}}Q &~=~& P\go{\omega}Q\\[-.5ex]
    P\go{\trace{a}t}Q &~=~& \exists P'\cdot
    P\go{a}P'~\land~P'\go{t}Q
\end{eqnarray*}
For a standard process $P$, the derived trace $DT(P)$ is defined as follows:
\begin{defn}\label{def1}
For a trace $t$,
  $ t\in DT(P) ~=~ P\go{t}0$
\end{defn}

For compensable processes, it is required to extract traces from
both forward and compensation behaviour. First, we define the
lifted forward behaviour and then add the behaviour of
compensation by reusing the above definition. For a compensable
process $PP$, we get the following definition:

\begin{defn}\label{def2}
For traces $t$ and $t'$,
\begin{eqnarray*}
 (t,t')\in DT(PP) &=& PP\go{(t,t')}0\\
 &=& \exists P'\cdot PP\go{t}P'\land P\go{t}0
\end{eqnarray*}
\end{defn}
Finally, the semantic relationship is defined as follows:
\begin{thm}\label{th1}
For a standard process term $P$ ($P\neq 0$),
$$DT(P)~=~T(P)$$
For a compensable process terms $PP$, where $PP\neq 0$
$$DT(PP) ~=~ T(PP)$$
\end{thm}
The theorem is proved by showing that
\begin{eqnarray*}
t\in DT(P)&~=~& t\in T(P)\\
(t,t')\in DT(PP)&~=~&(t,t')\in T(PP)
\end{eqnarray*}
We apply induction over process terms and define supporting lemmas for the structural cases. Traces are extracted for each term of the language and show their correspondence with the original trace semantics. For standard processes, $P$ and $Q$, for all the operators, we show that,
\begin{eqnarray}
t\in DT(P\otimes Q) &=& t\in T(P\otimes Q)\label{eq1}
\end{eqnarray}
For each such operator $\otimes$, the proof is performed by induction over traces assuming $DT(P)= T(P)$, and $DT(Q)=T(Q)$. For compensable processes, $PP$ and $QQ$, we show,
\begin{eqnarray}
(t,t')\in DT(PP\otimes QQ)~=~ (t,t')\in T(PP\otimes QQ)
\end{eqnarray}

Consider the sequential composition of processes $P$ and $Q$. By
using~(\ref{eq1}), the semantic relationship is shown by,
\begin{eqnarray*}
t\in DT(P\sq Q) &=& t\in T(P\sq Q)
\end{eqnarray*}
From Def.~\ref{def1}, we  get the following equation,
\begin{eqnarray*}
t\in DT(P\sq Q) &=& (P\sq Q)\go{t}0
\end{eqnarray*}
We also expand the definition of trace semantics as follows:
\begin{align*}
    t &\in~ T(P\seq Q)&&\\
      &=~\exists p,q\cdot t=(p\sq q)~~\land~~
        p\in T(P)~~\land~~q\in T(Q) &&\\
      &=~\exists p,q\cdot t=(p\sq q)~~\land~~
        p\in DT(P)~~\land~~q\in DT(Q) &&\\
      &=~\exists p,q\cdot t=(p\sq q)~~\land~~
      P\go{p}0~~\land~~Q\go{q}0&&
\end{align*}
Finally, from the above definitions of traces, the following lemma is formulated
for the sequential composition of standard processes:
\begin{lema}\label{lem:stseq}~\\
$(P\sq Q) \go{t}0 =\exists p,q\cdot t=(p\sq q)
\land P\go{p}0\land Q\go{q}0$
\end{lema}
The lemma is proved by applying induction over the trace $t$, where
$t=\trace{\omega}$ is the base case, and $t=\trace{a}t$ is
the inductive case. Similarly, the supporting lemmas for all the other terms of the language are defined and proved.

For synchronous processes, we follow the same approach added with the newly defined $\bot$ event. With the introduction of partial behaviour, the definition of derived traces remains the same except for the compensable processes. For a pair of traces ($t$ and $t'$), the derived traces of synchrnous compensable processes  is defined as follows:
$$PP\go{(t,t')}0 ~=~
\begin{cases}
 \exists R\cdot PP\go{t}R\land R\go{t'}0 & last(t)\neq\bot\\
  PP\go{t}0\land t'=\trace{\bot} & last(t)=\bot
\end{cases}$$

Considering Theorem~\ref{th1}, for synchronous processes we prove the following lemma:
\begin{lema}
For standard process terms $P$ and $Q$,
\begin{eqnarray*}
		DT(P\parallel_X Q) &=& T(P\parallel_X Q)
\end{eqnarray*}
For compensable process terms $PP$ and $QQ$,
\begin{eqnarray*}
		DT(PP\parallel_X QQ) &=& T(PP\parallel_X QQ)
\end{eqnarray*}
\end{lema}

By following the approach shown earlier we formulate the following lemma for standard processes:
\begin{lema}\label{lema:synstd}
$\begin{array}[t]{rl}
(P~\parallel_{X}~Q)\go{t}0 ~=&
      \exists p,q\cdot t\in(p~\parallel_{X}~q)\\
      & \land P\go{p}0~\land~Q\go{q}0
      \end{array}$
\end{lema}

Based on the scenario when synchronizing processes fail to synchronize and return partial behaviour, we state two separate lemmas.
First, we assume that there is no failure during the synchronization of processes:
\begin{lema}\label{lem:nondead}
$\begin{array}[t]{l}
 (PP\parallel_X QQ)\go{t}R ~=~\\[.5ex]
  \exists p,q,P,Q \cdot t\in (p\parallel_X q) \land last(t)\neq \bot\\[.5ex]
	\land  PP\go{p}P\land QQ\go{q}Q \land R~=~(P\parallel_X Q)
\end{array}$
\end{lema}

The following lemma is defined for the cases when the synchronizing
processes fail to synchronize:
\begin{lema}\label{lem:dead}~\\
$\begin{array}{ll}
(PP\parallel_X QQ)\go{t}0 ~=&
	\exists p,q \cdot t\in (p\parallel_X q)~\land~last(t)=\bot\\[.7ex]
	& \land  p\in T(PP) \land q\in T(QQ)
	\end{array}$
\end{lema}

In earlier work~\cite{entcs09}, we have shown how to mechanically proof the relationship between the asynchronous semantic models by embedding the cCSP syntax and semantic models into the theorem prover PVS, where the mechanical proofs have followed the similar proof steps as in hand proofs shown in~\cite{techreport-ecs}. After extending the semantic models to synchronization, instead of proving the relationship by hand, we directly prove them by using PVS. In the following section, we describe how we define and prove the semantic relationship for synchronous models by extending the asynchronous embeddings in PVS.

\section{Mechanizing Relationship}\label{sec:mechanize}

\begin{figure}[!htb]
\centering
\includegraphics[width=60 mm]{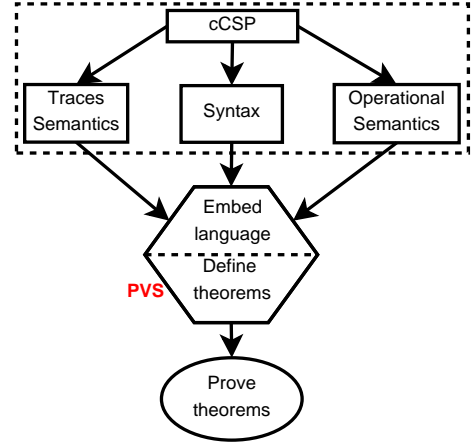}
\caption{PVS mechanization steps}
\label{fig:pvs-steps}
\end{figure}

An embedding is a semantic encoding of one specification language into another, especially, to reuse the existing tools of the target language. Mechanization steps of synchronous processes are outlined in this paper. Detail mechanization steps are described in~\cite{Ripon2008}. PVS mechanization steps are sketched in Fig.~\ref{fig:pvs-steps}.

\subsection{cCSP Syntax}

\begin{figure}[!htb]
\centering
\includegraphics[width= 84 mm]{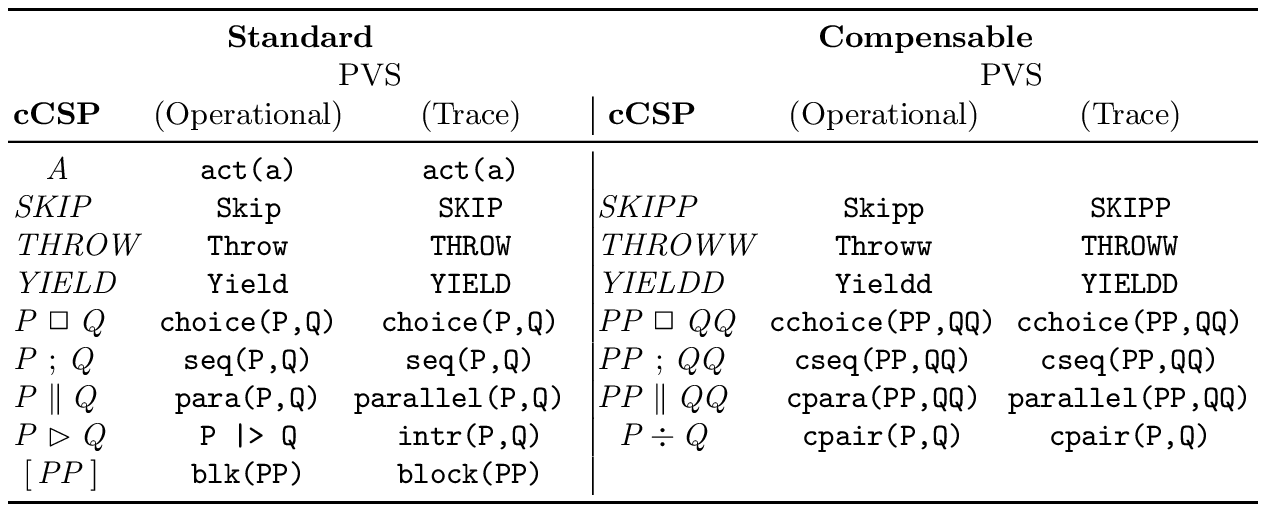}
\caption{cCSP syntax in PVS}
\label{fig:syntax-pvs}
\end{figure}

First, we define the cCSP syntax. Separate notation is used to define the standard and compensable processes. As PVS supports overloading, same notations can be used for the operational and the trace semantics. Fig.~\ref{fig:syntax-pvs} summarizes the PVS definition of asynchronous subset of cCSP syntax.

The syntax is then extended to define the terms for synchronization. To denote the trace semantics, we write \verb!full_parallel(X)(P,Q)! ($P\parallel_X Q$) for standard processes and \verb!cfull_parallel(X)(PP,QQ)! ($PP\parallel_X QQ$) for compensable processes.

\subsection{Process Algebra Terms}

Proofs about properties of a process algebra often use induction on the structure of the algebra. PVS has a mechanism called abstract datatype~\cite{datatype:PVS}, for which PVS generated an induction scheme, and it is convenient to model process algebra terms as an abstract datatype. cCSP has standard, and compensable process terms and importantly, these process terms are mutually dependant on each other. Mutually recursive datatype is not directly admissible by PVS. However, PVS has an extended support of \emph{sub-datatype}~\cite{datatype:PVS,subtype:PVS}, where it is possible to define two mutually recursive datatypes as a single datatype. A sub-datatype collects together groups of constructors of a datatype that form one part of a mutually recursive datatype definition. By using this facility we  define cCSP process algebra terms as follows:

\begin{script}
pa_terms  : DATATYPE WITH SUBTYPES stand, comp
 BEGIN
  Skip    : skip?   : stand
  choice(P: stand, Q: stand)     : choice?   : stand
  seq(P:stand, Q:stand)          : seq?      : stand
  |>(P: stand, Q: stand)         : inthnd?   : stand
  cseq(PP : comp, QQ : comp)     : c_seq?    : comp
  cchoice(PP : comp, QQ : comp)  : c_choice? : comp
  cpair(P: stand, Q : stand)     : cpair?    : comp
  blk(PP : comp)                 : blk?      : stand
  synpara(X:setof[normal],P:stand,Q:stand)
                                 :synpara?   : stand
  csynpara(X:setof[normal],PP:comp, QQ:comp)
                                 :csynpara?  : comp
  ...% other terms are omitted from this presentation
END pa_terms
\end{script}
\verb!synpara! and \verb!csynpara! are the extended definitions for the synchronous process terms. We define a single datatype \verb!pa_terms! that consists of two sub-datatypes: `\verb!stand!' for standard processes, and `\verb!comp!' for compensable processes. We can now define processes of types `\verb!stand!' and `\verb!comp!'.


\subsection{Trace Semantics}

The trace semantics are defined in PVS in the same way as they are originally defined. Operators are first defined at the trace level, and then lift to the sets of traces to define the processes. The same approach is taken for both standard, and compensable processes. For synchronous processes, we first define the synchronization of terminal evens shown in Table~\ref{tab:synch} by extending the asynchronous definition (\verb!parallel!).

\begin{script}
syn_parallel(w3:terminal)(w1,w2:terminal):bool=
 IF w3 = bottom THEN
    w1 = bottom OR w2 = bottom
 ELSE parallel(w3)(w1,w2) ENDIF
\end{script}
The trace semantics for synchronous processes are then defined by following the definitions shown in Sec.~\ref{sec:trace}. First we define operators over traces then lift it over set of traces to define processes. The trace semantics of both standard and compensable processes are defined in PVS as follows:

\begin{script}
full_parallel(X)((s1,w1))((s2,w2))((s3,w3)):RECURSIVE bool=
 CASES s3 OF
 null:null?(s1) AND null?(s2)  AND syn_parallel(w3)(w1,w2)
  OR cons?(s1) AND X(car(s1)) AND null?(s2) AND w3 = bottom
  OR cons?(s2) AND X(car(s2)) AND null?(s1) AND w3 = bottom
  OR cons?(s1) AND X(car(s1)) AND cons?(s2) AND X(car(s2))
              AND  car(s1) /= car(s2) AND w3 = bottom,
 cons(a,tail):
  IF X(a) THEN cons?(s1) AND cons?(s2)     AND
               car(s1) = a AND car(s2) = a AND
    full_parallel(X)((cdr(s1),w1))((cdr(s2),w2))((tail,w3))
  ELSE cons?(s1) AND car(s1) = a AND
        full_parallel(X)((cdr(s1),w1))((s2,w2))((tail,w3))
     OR cons?(s2) AND car(s2) = a AND
        full_parallel(X)((s1,w1))((cdr(s2),w2))((tail,w3))
  ENDIF ENDCASES
 MEASURE length(s3)
full_parallel(X)(P,Q : process): process =
{t : trace | EXISTS (p:(P),q:(Q),s1,w1,s2,w2,s3,w3):
		    p = (s1,w1) AND q = (s2,w2) AND t = (s3,w3) AND
		    full_parallel(X)((s1,w1))((s2,w2))((s3,w3)) }

cfull_parallel(X)((p,p1))((q,q1))((r,r1)) : bool =
    (full_parallel(X)(p)(q)(r) AND
    full_parallel(X)(p1)(q1)(r1) AND r`2 /= bottom)
 OR full_parallel(X)(p)(q)(r) AND
    r`2 = bottom AND null?(r1`1) AND r1`2 = bottom
 cfull_parallel(X)(PP,QQ:comp_process):comp_process=
 { tt:comp_trace | EXISTS (pp:(PP),qq:(QQ)) :
                  cfull_parallel(X)(pp)(qq)(tt) }
\end{script}
We represent traces as a pair: \verb!(s,w)!, where \verb!s! is the sequence of normal events and \verb!w! is the terminal event.

\subsection{Operational Semantics}

The operational semantics is defined by using labelled transition systems of the form $P\go{e}P'$, where the event $e$ makes the transition of the process term from state $P$ to $P'$. Two types of transitions are defined: normal, and terminal. Both transition rules are defined by using a recursive boolean definition that determines whether there is a transition from one state to another state. The definitions are given by using equations derived from the transition rules.  The transition rules of some process terms depend on the transition rules of both standard and compensable processes. To define these rules, we need to combine the transition rules for both standard and compensable processes. The terminal transition for the process terms are defines as \verb!wtrans! and the normal transitions are defined as \verb!ntrans! (See~\cite{Ripon2008},\cite{entcs09} for details). We then define the transition rules for synchronous processes by following the definitions given in Fig.~\ref{fig:synch-op} and \ref{fig:csyn-op}.

In a normal transition, processes either synchronize or interleave. By extending the transition rules of asynchronous processes we defne the transition rules for synchronous processes as follows:

\begin{script}
synpara(X,Q,R):
 IF X(a) THEN
  EXISTS Q1,R1 : ntrans(a)(Q,Q1) AND  ntrans(a)(R,R1) AND
                Pa1 = synpara(X,Q1,R1)
 ELSE EXISTS Q1: ntrans(a)(Q,Q1) AND Pa1 = synpara(X,Q1,R)
   OR EXISTS R1: ntrans(a)(R,R1) AND Pa1 = synpara(X,Q,R1)
 ENDIF
csynpara(X,QQ,RR) :
IF X(a) THEN
 EXISTS QQ1,RR1:ntrans(a)(QQ,QQ1) AND ntrans(a)(RR,RR1) AND
                 Pa1 = csynpara(X,QQ1,RR1)
 ELSE
 EXISTS QQ1:ntrans(a)(QQ,QQ1) AND Pa1 = csynpara(X,QQ1,RR)
 OR EXISTS RR1:ntrans(a)(RR,RR1) AND Pa1= csynpara(X,QQ,RR1)
\end{script}

The terminal transitions are defined as follows:

\begin{script}
synpara(X,Q,R):
    EXISTS w1,w2: syn_wtrans(w1)(Q,nul) AND
                  syn_wtrans(w2)(R,nul) AND
                  syn_parallel(w)(w1,w2) AND P1 = nul
 OR EXISTS (a:normal,w1,Q1): X(a) AND ntrans(a)(Q,Q1) AND
     syn_wtrans(w1)(R,nul) AND w = bottom AND P1 = nul
 OR EXISTS (a:normal,w1,R1) : X(a) AND ntrans(a)(R,R1) AND
     syn_wtrans(w1)(Q,nul) AND w = bottom AND P1 = nul
 OR EXISTS (a1,a2:normal,Q1,R1):
                X(a1) AND X(a2) AND a1 /= a2 AND
                ntrans(a1)(Q,Q1) AND ntrans(a2)(R,R1) AND
                 w = bottom AND P1 = nul,
csynpara(X,QQ,RR):
    EXISTS Q1,R1,w1,w2 : syn_wtrans(w1)(QQ,Q1) AND
        syn_wtrans(w2)(RR,R1) AND syn_parallel(w)(w1,w2)
        AND w /= bottom AND P1 = synpara(X,Q1,R1)
 OR EXISTS (a:normal,w1,QQ1,R1): X(a) AND
         ntrans(a)(QQ,QQ1) AND syn_wtrans(w1)(RR,R1) AND
         w = bottom AND P1= nul	
 OR EXISTS (a:normal,w1,Q1,RR1): X(a) AND
         syn_wtrans(w1)(QQ,Q1) AND ntrans(a)(RR,RR1) AND
         w = bottom and P1 = nul
 OR EXISTS (a1,a2:normal,QQ1,RR1):
        X(a1) AND X(a2) AND a1 /= a2 AND
        ntrans(a1)(QQ,QQ1) AND ntrans(a2)(RR,RR1) AND
        w = bottom AND P1 = nul
\end{script}

\subsection{Semantic Relationship}

By following Def.~\ref{def1}, the derived traces for standard processes are defined as `\verb!trans_trace!'. It defines the transition of a process by a trace consisting of a transition by a sequence of normal events followed by transition by a terminal event. Consider a trace $t$, where $t=t'\trace{\omega}$.
\begin{eqnarray*}
P\go{t'\trace{\omega}}0&=& \exists P'\cdot P\go{t'}P'\land P'\go{\omega}0
\end{eqnarray*}

We then define Lemma~\ref{lema:synstd} by using the definition of both derived traces and trace rules as follows:

\begin{scriptsize}
\begin{Verbatim}[frame=single]
synpara_lemma : LEMMA
  trans_trace((s,w))(synpara(X,P,Q),nul) =
   EXISTS (s1,w1,s2,w2) :
    full_parallel(X)((s1,w1))((s2,w2))((s,w)) AND
    trans_trace((s1,w1))(P,nul)               AND
    trans_trace((s2,w2))(Q,nul)
\end{Verbatim}
\end{scriptsize}

For compensable processes, we only need to prove that the lifted forward behaviour corresponds to the original traces and reuse the proofs of standard processes for compensations. The definition of derived traces shown in Def.~\ref{def2} consists of the derived trace of both forward and compensation behaviour. To prove our lemmas (Lemma~\ref{lem:nondead} and \ref{lem:dead}) we only need to define the forward behaviour and it is defined as \verb!ftrans_trace! ($PP\go{t}P$).

First, we define the lemma considering the processes will not fail to synchronize and hence, there is no bottom event in the derived traces:

\begin{scriptsize}
\begin{Verbatim}[frame=single]
csynpara_lemma : LEMMA
 ftrans_trace((s,w))(csynpara(X,PP,QQ),R) =
   EXISTS (s1,w1,s2,w2,P,Q): w /= bottom AND
   full_parallel(X)((s1,w1))((s2,w2))((s,w)) AND
   ftrans_trace((s1,w1))(PP,P) AND
   ftrans_trace((s2,w2))(QQ,Q) AND
   R = synpara(X,P,Q)
\end{Verbatim}
\end{scriptsize}

Next, we define the lemma where compensable processes fail to synchronize during their synchronization. The main difference is that the derived trace now ends with a $\bot$ representing the partial behaviour, and compensations are not accumulated after termination.

\begin{scriptsize}
\begin{Verbatim}[frame=single]
lema_bot : LEMMA
 ftrans_trace((s,w))(csynpara(X,PP,QQ),nul) =
  EXISTS (s1,w1,s2,w2,P,Q):  w = bottom AND
   full_parallel(X)((s1,w1))((s2,w2))((s,w)) AND
   ftrans_trace((s1,w1))(PP,P)               AND
   ftrans_trace((s2,w2))(QQ,Q)
\end{Verbatim}
\end{scriptsize}

All these lemmas are proved interactively by applying induction over traces (\verb!(s,w)!). PVS has a strong support for induction scheme which facilities proving such lemmas.

\section{Related Work}

One of the contributions most related to our work is by Basten and Hooman in~\cite{papvs}, where the focus is  on the use of a general purpose proof checker, e.g., tool support for the proof of theoretical properties of an ACP-style process algebra~\cite{acp} . The idea is to apply equational reasoning. Mechanical support for both verification of concrete applications and proving theoretical properties of the process algebra are investigated.

PVS has been used in~\cite{trace-pvs,neil:pvstrace} to mechanize the trace semantics of CSP. Their goal is to verify an authentication protocol specified in CSP to overcome errors in the manual verification as well as improve the scalability of the approach. The mechanization is
based on a semantic embedding of CSP. The traces are defined by using a list of events and processes are defined by prefix-closed sets of traces. The important distinction with the present work is that cCSP traces are non-empty and completed and processes are defined accordingly.

Camilleri \cite{csphol} showed how to mechanize a subset of the CSP operators by using the theorem prover HOL \cite{HOL}. The trace model for a subset of the CSP operators was mechanized in HOL. Initially, events, alphabets and traces are defined and then CSP operators are defined in terms of their trace semantic models. And later laws related to the operators are proved from the sematic definition. In contrast to our approach no syntax is defined at this stage and operators are defined directly in HOL. Syntax is defined later and the semantics of the language is shown based on the already defined semantics. A similar work for the $\pi$-calculus can be found in~\cite{pihol}. One of our main goals is to explore the ways of incorporating process algebra in a general purpose theorem prover. In that respect, a closely related research on the tool support for a process algebra shown in \cite{di-hol}, where a CSP-like algebra, called DI-Algebra \cite{DI-algebra} is formalized in HOL. The
algebra is used to reason about synchronous circuits. Process syntax and algebraic laws are defined, but no semantics are defined.

\section{Concluding Remarks}\label{sec:conclusion}

We have extended cCSP language to define synchronization. We introduced the notion of partial behaviour which allows to model the behaviour of synchronous processes that fail to synchronize. The formal foundation of the language is strengthen by establishing a relationship between the semantic models by showing that traces extracted from the operational semantics correspond to the original trace semantics. Demonstrating the relationship between these two semantics of the ensures the consistency of the semantic
description of the language.

We have started mechanizing the semantic models and their relationship in order to investigate the feasibility of the mechanization process. We have achieved our goal by successfully proving the semantic relationship for the synchronous processes. Defining process algebras in PVS is not new a new idea. The novelty of this experiment is that, we have not only defined the cCSP process algebra, and the two semantic models, but we have also mechanically proved a relationship between these semantic models.

In the hand proofs, it is easy to be imprecise about recursion, and typing of the rules. The mechanization forces to be strict about datatypes, and recursion. This helped us to define the theorems, and the lemmas in a systematic way, and to prove all the lemmas by following a similar fashion. The mechanization also helped us identifying some lemmas which were not explored earlier. The mechanization of the semantic models and their relationships also deepen our understanding of the semantic models for both standard and compensable processes.

Having a firm grasp of the semantic models, we are now in a better position to extend the language by defining some important operators for the process algebra, such as event hiding, recursion, distinction between external and internal choice in combination with compensations. In standard CSP, the distinction between the two choice operators is achieved by using the Failure/Divergences model which can serve as the basis for our work on cCSP. Our future plan also includes developing a tool support for cCSP which will allow model check as well as animate the specifications.

\section*{Acknowledgement}
The research has been carried out at the University of Southampton, as a part of the first author's PhD project.

\IEEEtriggeratref{11}

\end{document}